\documentclass[aps,prl,twocolumn,groupedaddress,showpacs,floatfix]{revtex4-1}
\usepackage{graphicx}
\usepackage{amsmath}
\usepackage{mathrsfs}
\usepackage{amsmath}
\usepackage{amsfonts}
\usepackage{amssymb}
\usepackage{makeidx}
\usepackage[section] {placeins}
\newcommand{\ket}[1]{{\left| {#1} \right\rangle}}
\newcommand{\bra}[1]{{\left\langle {#1} \right|}}

\begin{document}
\title{Coherent Error Suppression in Multi-Qubit Entangling Gates}
\author{D. Hayes}
\email[dhayes12@umd.edu]{}
\author{S.~M. Clark}
\author{S. Debnath}
\author{D. Hucul}
\author{I.~V. Inlek}
\author{K.~W. Lee}
\author{Q. Quraishi}
\author{C. Monroe}
\affiliation{Joint Quantum Institute and Department of Physics,
University of Maryland, College Park, MD 20742, USA}

\begin{abstract}
We demonstrate a simple pulse shaping technique designed to improve the fidelity of spin-dependent force operations commonly used to implement entangling gates in trapped-ion systems.  This extension of the M{\o}lmer-S{\o}rensen gate can theoretically suppress the effects of certain frequency and timing errors to any desired order and is demonstrated through Walsh modulation of a two-qubit entangling gate on trapped atomic ions.  The technique is applicable to any system of qubits coupled through collective harmonic oscillator modes.
\end{abstract}
\date{\today}
\maketitle

The use of spin-dependent forces to create entangled quantum systems has become widespread \cite{sackett:2000,blatt:2009,Lee:2003,steane:2006} and is currently the technique used for the highest fidelity multi-qubit operations \cite{blatt:highfidelity:2008}.  This powerful technique, first proposed in \cite{molmer:sorensen:1999,james:2000,solano:1999}, has been used to implement quantum algorithms \cite{blatt:2008}, create large entangled states \cite{blatt:2010}, test quantum fundamentals \cite{rowe:2000,blatt:kochenspecker:2009}, and perform simulations of quantum magnetism \cite{monroe:magnetism:2010,Friedenauer:2008} and quantum field theory \cite{blatt:zitterbewegung:2010}.  As these types of experiments are scaled to larger numbers of qubits and more complex algorithms, the accumulation of gate errors will eventually require quantum error correction.  Because of the large overhead required for quantum error correction, it is important that qubit operations be optimized passively in terms of speed and robustness to non-ideal control environments.   In this paper, we show how ideas similar to the spin-echo pulse sequence \cite{Carr:1954} and those developed in the context of dynamical decoupling \cite{viola:1998,uhrig:2007,biercuk:2009} can be used to optimize the M{\o}lmer-S{\o}rensen (MS) gate that is based on the spin-dependent force.

Spin-dependent force gates operate by coupling the qubit states to a collective external degree of freedom referred to as a quantum bus.  The coupling is switched on for an amount of time that introduces a particular phase between the spin states and at the same time leaves them disentangled from the external degree of freedom at the end of the gate.  While the relative spin phase is fairly robust due to geometric features \cite{zhu:2003}, the disentanglement of the qubit space and the quantum bus at the end of the operation may be more susceptible to experimental errors and is equally crucial to achieving a high fidelity gate.  Imperfect timing caused by noise on the energy splitting of the qubit can be suppressed by the insertion of an additional swapping pulse on the qubit states in the middle of a two qubit gate \cite{Jost:2009} or, as proposed in \cite{chen:2008}, by a $\pi$ phase shift in the drive field.  In this Letter, we present new analytic results which generalize these ideas and show how frequency and timing errors can theoretically be suppressed to any desired order with an optimized gate sequence that does not rely on the insertion of additional $\pi$ pulses within the gate.  Furthermore, the technique is demonstrated using atomic hyperfine qubits driven by a stimulated Raman process and shown to perform much better than the standard operation described in the original proposal \cite{molmer:sorensen:1999}.  Similar to the single-qubit composite pulses \cite{Cummins:2003} originally designed for error suppression in NMR experiments now being in widespread use in other quantum information systems, this composite pulse should be applicable to any system of qubits coupled to a driven harmonic oscillator such as superconducting flux qubits \cite{Wang:2009} or cavity QED \cite{chen:2008}.

In trapped-ion systems, the spin-dependent force couples internal atomic states of neighboring ions through the collective modes of motion generated by the Coulomb interaction.  In the MS scheme, a spin-dependent force is created by off-resonantly driving the first-order red and blue sideband transitions simultaneously.  The interaction Hamiltonian takes the form $\hat{H}=\Omega/2(\hat{\sigma}_+e^{i\phi_s}+\hat{\sigma}_-e^{-i\phi_s})(\hat{a}e^{-i\delta t}e^{i\phi_m}+\hat{a}^{\dagger}e^{i\delta t}e^{-i\phi_m})$ where $\Omega$ is the sideband transition frequency, $\hat{\sigma}_{\pm}$ are the raising and lowering operators for the qubit, $\left\{\hat{a}^{\dagger},\hat{a}\right\}$ are the creation and annihilation operators for the collective harmonic oscillator mode, and $\delta/2\pi$ is the symmetric detuning of the drive field from the sidebands \cite{molmer:sorensen:1999}.  The sum phase $\phi_s=(\phi_b+\phi_r)/2$ of the red and blue sideband drive fields determines the eigenstates of the spin operator in $\hat{H}$.  The difference phase $\phi_m=(\phi_b-\phi_r)/2$ determines the phase of the time-dependent displacement of the motional state.  For the general case of $N$ ions, the time-evolution operator is given by,
\begin{eqnarray}
\hat{U}(t)&=&e^{-i\int_{0}^tdt'\hat{H}(t')-\frac{1}{2}\int_{0}^{t}dt'\int_{0}^{t'}dt''\left[\hat{H}(t'),\hat{H}(t'')\right]}\\
&=&e^{\hat{S}_N(\alpha(t)\hat{a}^{\dagger}-\alpha^*(t)\hat{a})}e^{-i\Phi(t)\hat{S}_N^2},
\label{eq:time:evolution}
\end{eqnarray}
where the total spin operator is given by $\hat{S}_N=\sum_{i=1}^N\sigma^{(i)}_+e^{i\phi_s}+\sigma^{(i)}_-e^{-i\phi_s}$, the time-dependent displacement coefficient is $\alpha(t)=\Omega/2\int_{0}^{t}dt' e^{-i(\delta)t'}e^{i\phi_m}$ and $\Phi(t)$ is a time-dependent phase that depends only on $\Omega$ and $\delta$.  When a collection of trapped ions that are each identically prepared in an eigenstate of $\hat{\sigma}_z$ evolves according to (\ref{eq:time:evolution}), the spin-dependent displacement operator splits the motional wavepacket into $N+1$ pieces that execute circular trajectories in phase space according to the definition of $\alpha(t)$.  The term in (\ref{eq:time:evolution}) proportional to $\hat{S}_N^2$ imprints a relative phase on the various spin states, allowing the operation to be used as an entangling operation.  

\begin{figure}[ht]
\includegraphics[width=3.4in]{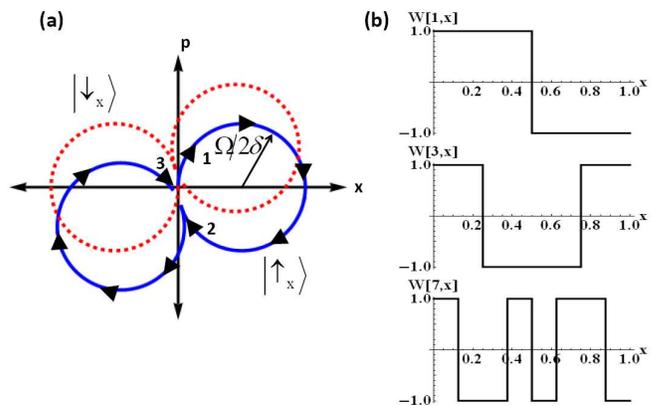}
\caption{(a)The phase space trajectory of the motional wavepackets of a single ion during a $\mathrm{W}(1,t/t_g)$ spin-dependent force operation with a small detuning error, $\Delta/\delta\ll1$.  The solid and dashed curves show the two different trajectories taken by the two different wavepackets associated with spin up and spin down in the $\hat{S}_{1}$ basis with only the spin up trajectory being labeled for clarity.  After being initialized to a state centered at the origin, the spin-up wave-packet begins its clockwise motion near the point labeled $1$.  Halfway through the operation, near the point $2$, the phase of the drive field is advanced by $\pi$, changing the direction of the applied force.  At the end of the gate, near point $3$, the wave-packet ends up much closer to the origin than the turning point near point 2.  Therefore, the two wave-packets have more overlap at the end of the gate which is the key to achieving a higher fidelity operation. (b)The Walsh functions $\mathrm{W}\left(1,x\right)$, $\mathrm{W}\left(3,x\right)$ and $\mathrm{W}\left(7,x\right)$ are shown.  Notice that $\mathrm{W}(3,x)$ can be constructed as two sequential $\mathrm{W}(1,x)$ functions with a phase flip on the second pulse.  Likewise, $\mathrm{W}(7,x)$ can be constructed as sequential $\mathrm{W}(3,x)$ pulses with a phase flip on the second pulse.}
\label{fig:phase:space}
\end{figure}

In order to prepare a pure spin state with this type of operation, the entanglement between the spin and motion must disappear at the end of the gate.  When the gate time $t_g$ is not equal to $2\pi j/\delta$, where $j$ is any non-zero integer, the motional wavepackets do not trace out closed trajectories in phase space and therefore will not be fully disentangled from the spin state.  The required level of precision grows with higher temperatures since the overlap between two states separated in phase space decreases exponentially with temperature.  To see this, consider a qubit under the influence of the time evolution operator in (\ref{eq:time:evolution}).  If the initial motional state is assumed to be a Gaussian state $\psi(x)$ with an uncertainty in position $\Delta x$ and we describe a small timing or detuning error in the gate operation as an unintentional momentum displacement $\hbar q$, then the overlap between the two motional states is given by $\int_{-\infty}^{\infty}dx\psi^*(x)e^{-iqx}\psi(x)=\mathrm{exp}\left[-\frac{1}{2}(q\Delta x)^2\right]$.  For a harmonic oscillator in a thermal state, $\Delta x$ increases approximately as $\sqrt{T}$ for $k_BT>\hbar\omega$ meaning that the overlap between the two states decreases exponentially.  As shown in Fig. \ref{fig:phase:space}(a), a small detuning error can largely be corrected with a second pulse whose phase has been shifted by $\pi$.  We now discuss how to generalize this simple pulse sequence in order to suppress larger errors of this type.

\begin{figure}[ht]
\includegraphics[width=1.6in]{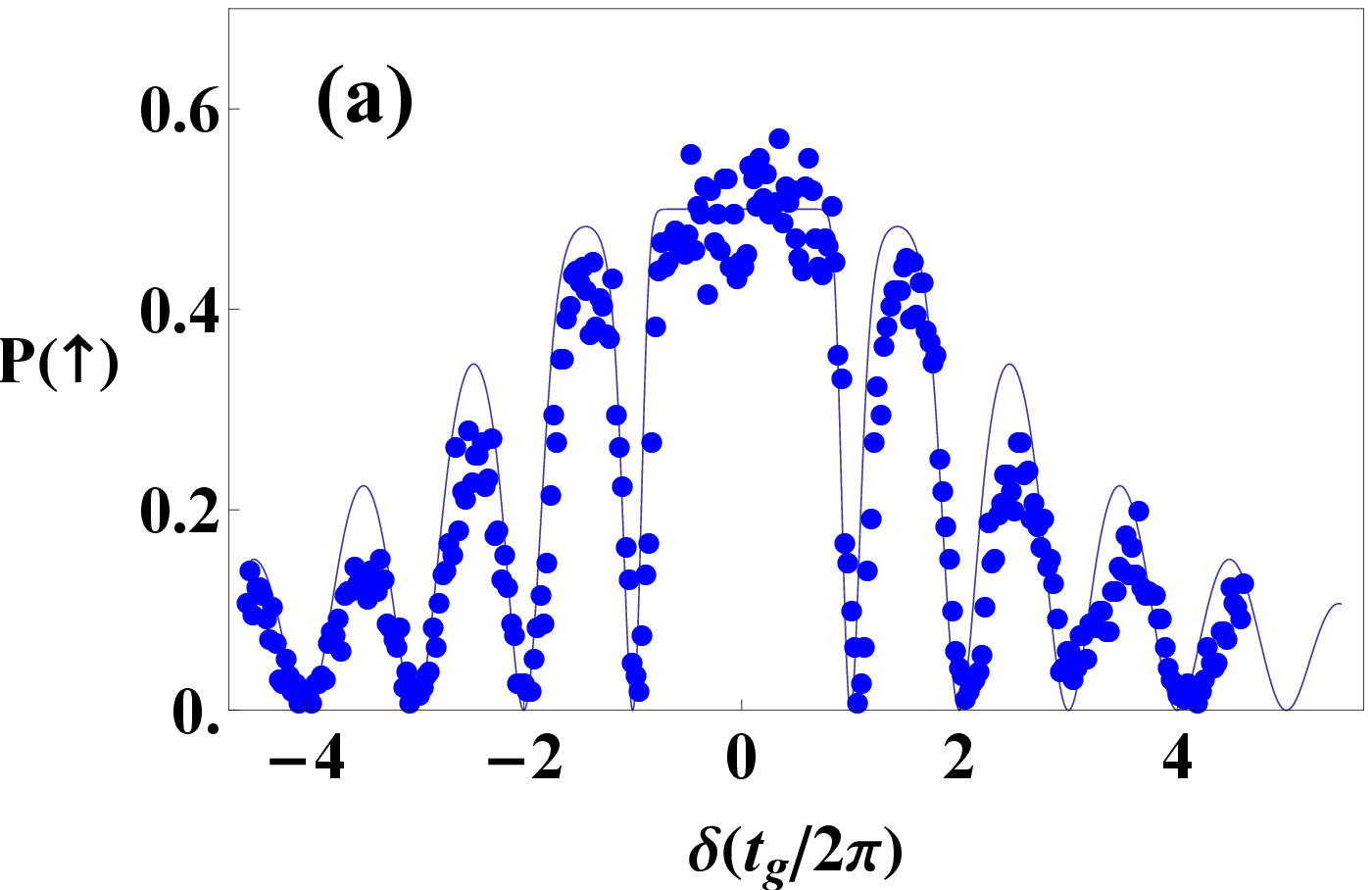}
\includegraphics[width=1.6in]{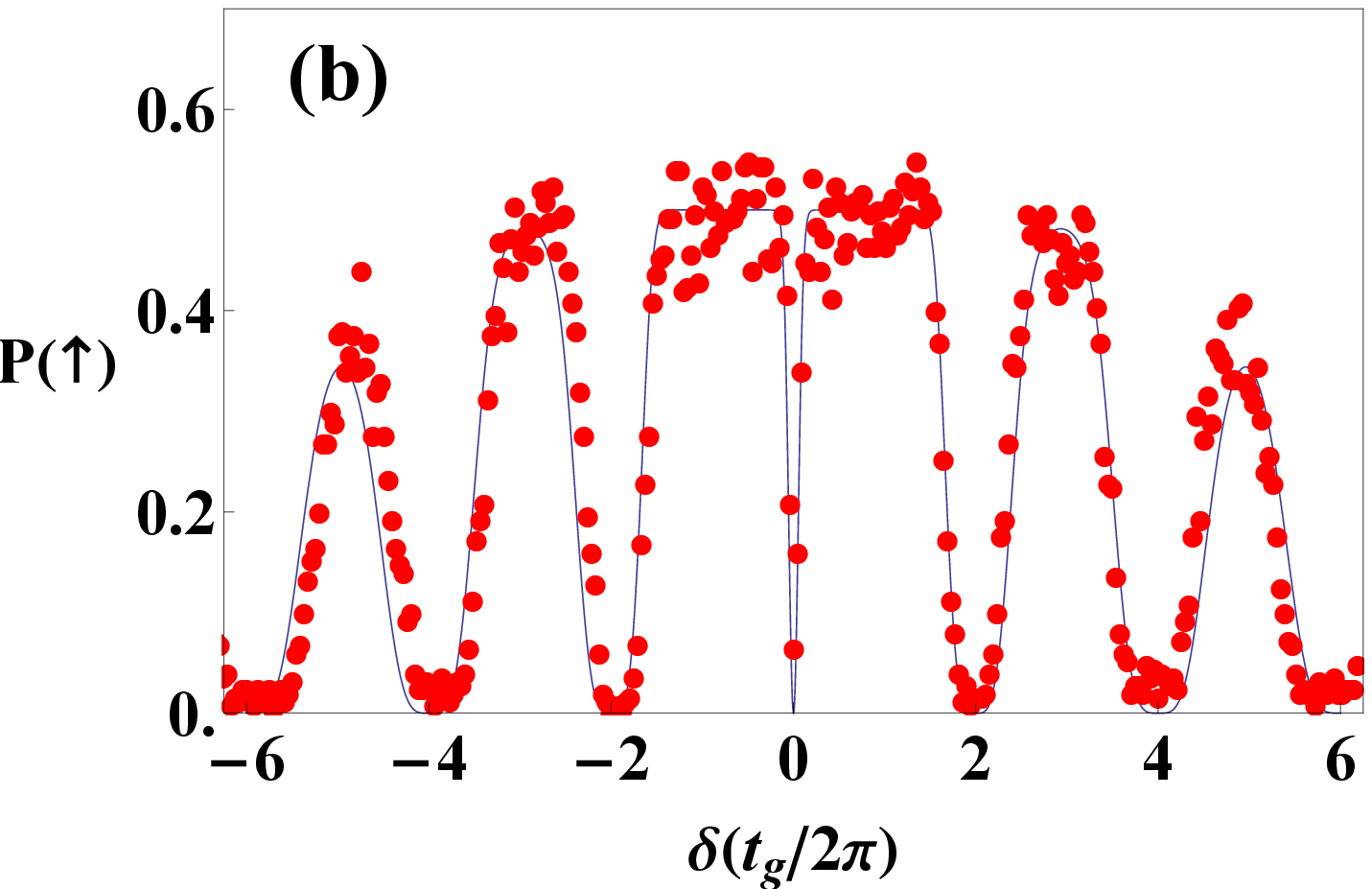}
\includegraphics[width=1.6in]{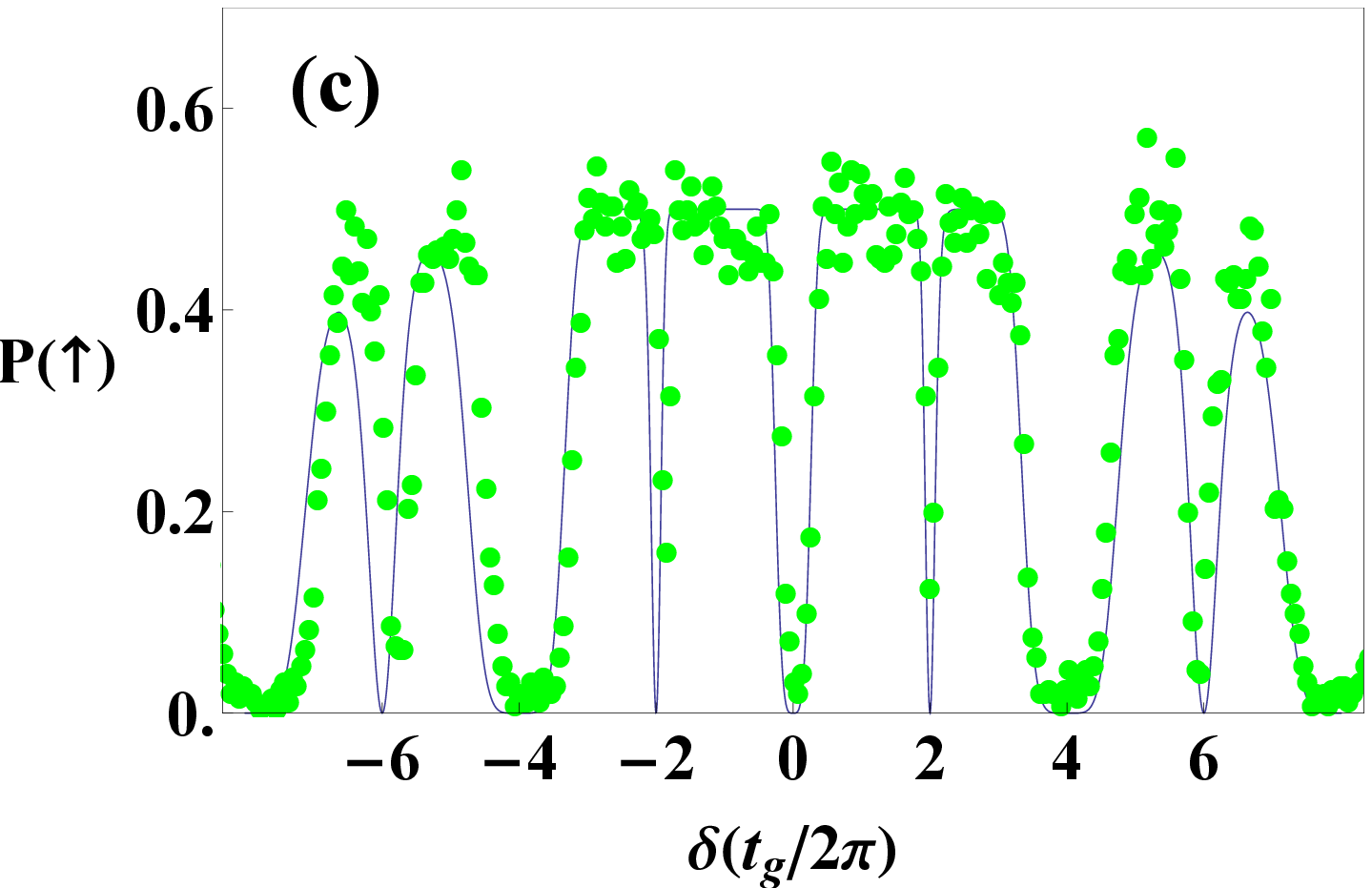}
\caption{A single ion prepared in $\ket{\downarrow}$ is subjected to the standard and composite spin-dependent force operations and then measured in the $\hat{\sigma}_z$ basis.  The data shown are plotted together with theoretical curves assuming an initial thermal state of motion with the average excitation number $\bar{n}=7$.  (a)  The data show the probability of finding the ion in $\ket{\uparrow}$ as a function of the symmetric detuning $\delta$ for $t_g=100~\mu$sec.  On resonance, $\delta=0$, the motional wave-packets quickly become entangled with the spin state, resulting in a maximally mixed spin state.  For finite $\delta$, the wavepackets trace out circles in phase space resulting in partial revivals of the initial spin state when $\delta t_g/2\pi$ is a non-zero integer.  (b)  The spin-dependent force operation is implemented using $\mathrm{W}(1,t/\sqrt{2}t_g)$ for the phase $\phi_s$.  (c)  $\mathrm{W}(3,t/2t_g)$ is used for the phase $\phi_s$.  Note the narrow resonance at $\delta t_g/2\pi=2$ corresponds to a trajectory where the phase flips occur when the motional wavepackets are not at the origin.}
\label{fig:one:ion}
\end{figure}
Suppose there is a symmetric error $\Delta$ in the detuning such that $\delta=2\pi/t_g+\Delta$ that could be the result of a change in the trapping frequency.  The error in the operation results in some residual entanglement between the spin and motion that can be quantified by the magnitude of $\alpha_0(t_g)=\Omega/2\int_{0}^{t_g}dt e^{-i(\delta+\Delta)t}$, (which goes to zero for $\Delta=0$ at $t_g=2\pi j/\delta$).  We will show that by switching either $\phi_s$ or $\phi_m$ between $0$ and $\pi$ at times prescribed by certain Walsh functions, the effect of $\Delta$ on the magnitude of $\alpha(t_g)$ can be suppressed to any order.  A Walsh function, denoted here as $\mathrm{W}\left(k,x\right)$, is a piecewise constant function that alternates between the values $\pm1$ at certain values of $x$ depending on the dyadic-ordered index $k$ \cite{beauchamp:1984}, (see Fig. \ref{fig:phase:space}(b)).  If $\phi_r$ and $\phi_b$ shift together between $0$ and $\pi$, then $\phi_s$ shifts between $0$ and $\pi$, but $\phi_m$ remains constant and can be assumed to be $0$ without the loss of generality.  Note the effect of the phase shift $\phi_s=0\Rightarrow\phi_s=\pi$ is equivalent to shifting the motional phase $\phi_m=0\Rightarrow\phi_m=\pi$ while keeping $\phi_s$ constant.  Both of these phase shifts are equivalent to the mapping $\hat{H}\Rightarrow-\hat{H}$, which can also be achieved with $\pi$ pulses on the qubit states as done in \cite{Jost:2009}.  Although the $\pi$ phase shifts and $\pi$ rotations are ideally equivalent, the phase shift switching time and precision is limited by electronics whereas the microwave rotations depend on qubit control that might be subject to the same noise source that generates $\Delta$.  If the times at which phase shifts occur is determined by the zero crossing times of $\mathrm{W}\left(k,t/t_g\right)$, then $\hat{S}_N=\mathrm{W}(k,t/t_g)\sum_i^N\sigma^{(i)}_++\sigma^{(i)}_-\equiv\mathrm{W}(k,t/t_g)\hat{X}_N$.  When modulating $\phi_s$ in this manner, the new displacement operator in (\ref{eq:time:evolution}) is given by,
\begin{equation}
\hat{D}_k(t_g)=e^{-i\hat{X}_N\frac{\Omega}{2}\int_{0}^{t_g}dt\mathrm{W(k,t/t_g)}(e^{-i(\delta+\Delta)t}\hat{a}^{\dagger}+e^{i(\delta+\Delta)t}\hat{a})}.
\label{eq:pulse:shaped:evolution}
\end{equation}
By choosing a Walsh function with index $k=2^n-1$ where $n$ is an integer and a detuning $\delta=2^{n+1}\pi/t_g$, phase flips only occur at integer multiples of $2\pi/\delta$ and the effect of $\Delta$ can be suppressed to any order.  This statement rests on the following equality,
\begin{equation}
\int_0^1dx\mathrm{W}(2^n-1,x)e^{\pm i2^{n+1}\pi x}\sum_{l=0}^na_lx^l=0,
\label{eq:walsh:coefficients}
\end{equation}
where $a_l$ is a constant, (proof in the supplemental material).  If the function $e^{\pm i\Delta t}$ is expanded in a Taylor series, the identity in (\ref{eq:walsh:coefficients}) ensures that the displacement operator will be given by $\hat{D}_{2^n-1}(t_g)=\hat{\mathrm{I}}+\mathcal{O}(\Delta^{n+1})$, where $\hat{\mathrm{I}}$ is the identity operator.
\begin{figure}[ht]
\includegraphics[width=1.6in]{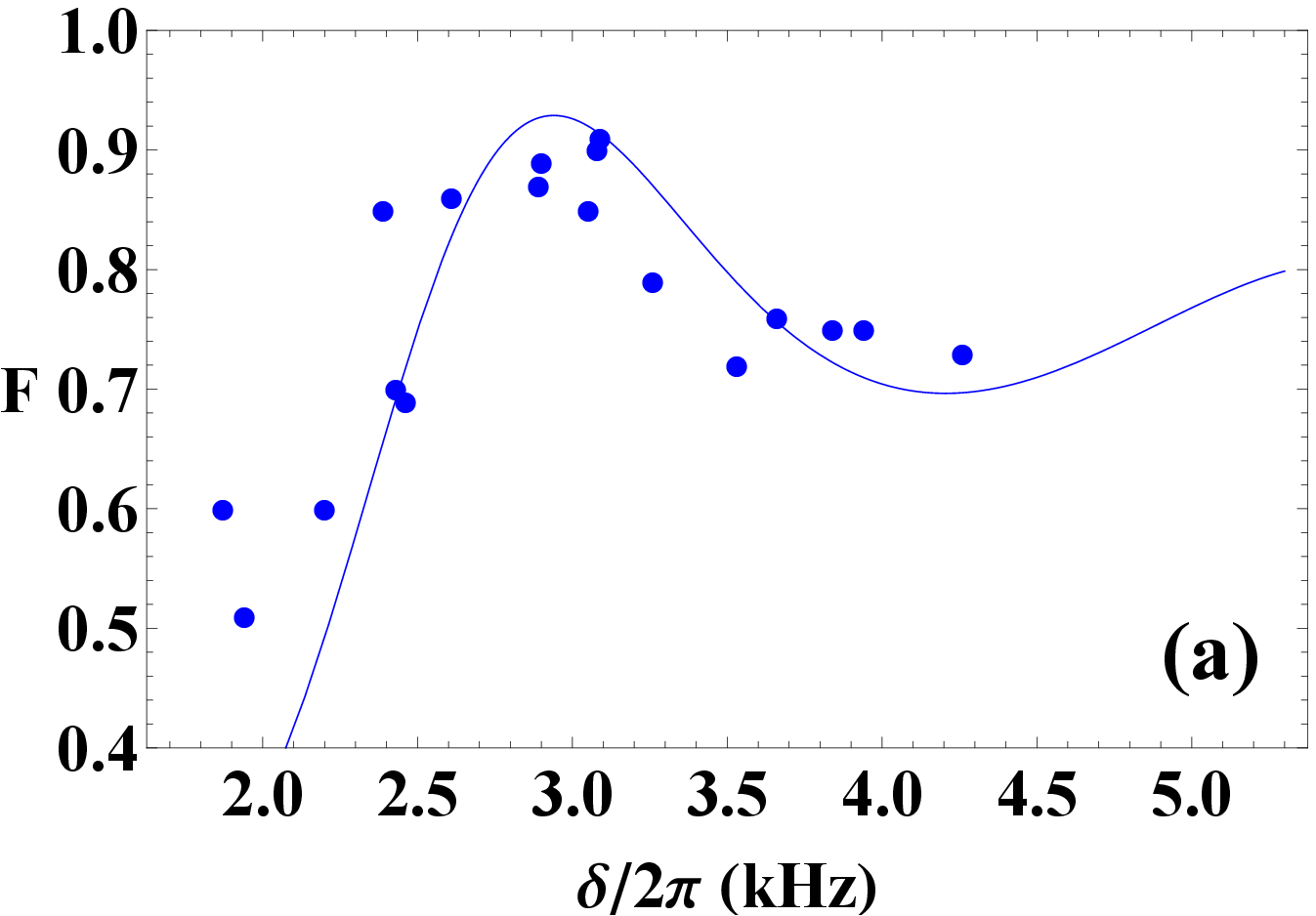}
\includegraphics[width=1.6in]{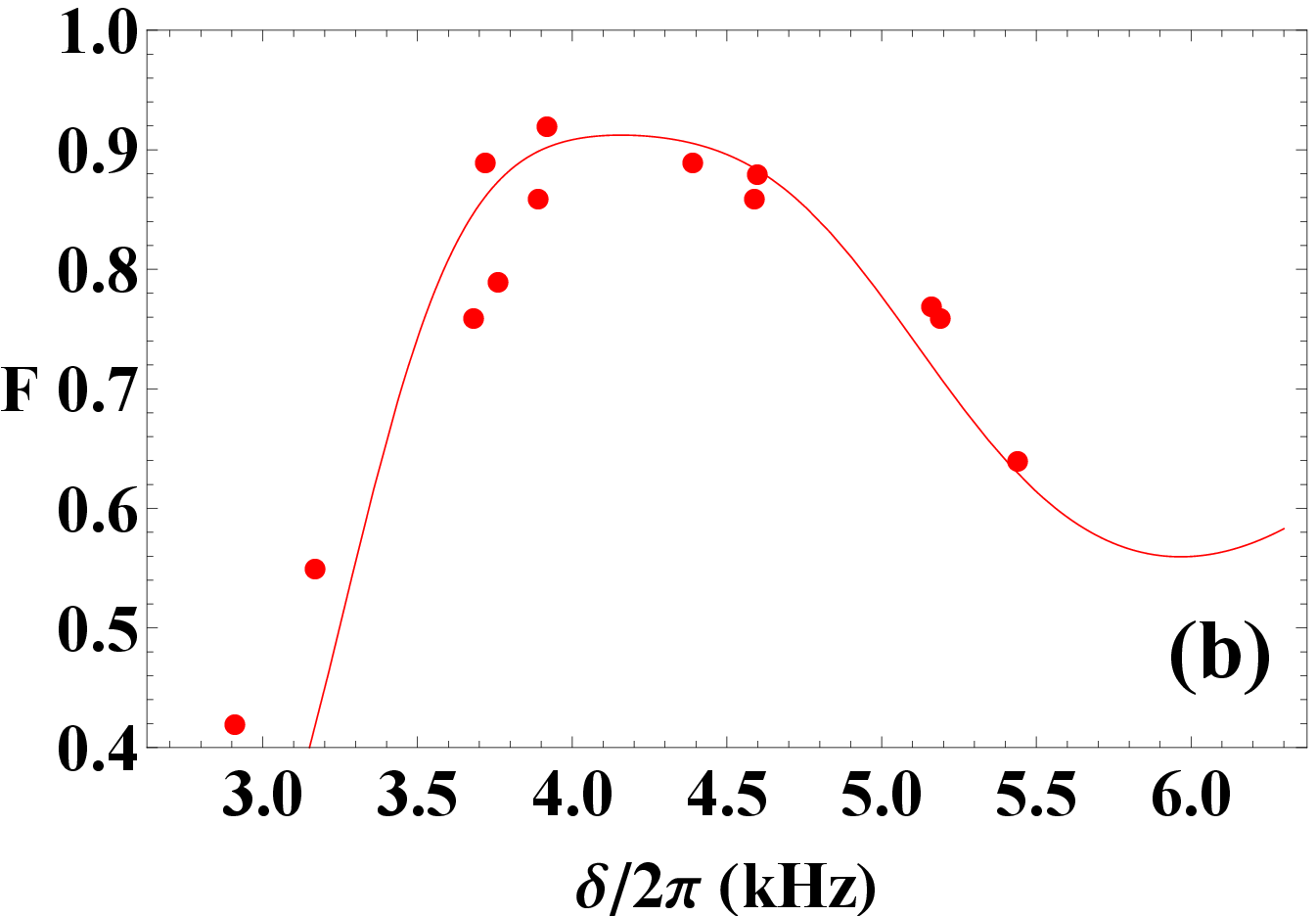}
\includegraphics[width=1.6in]{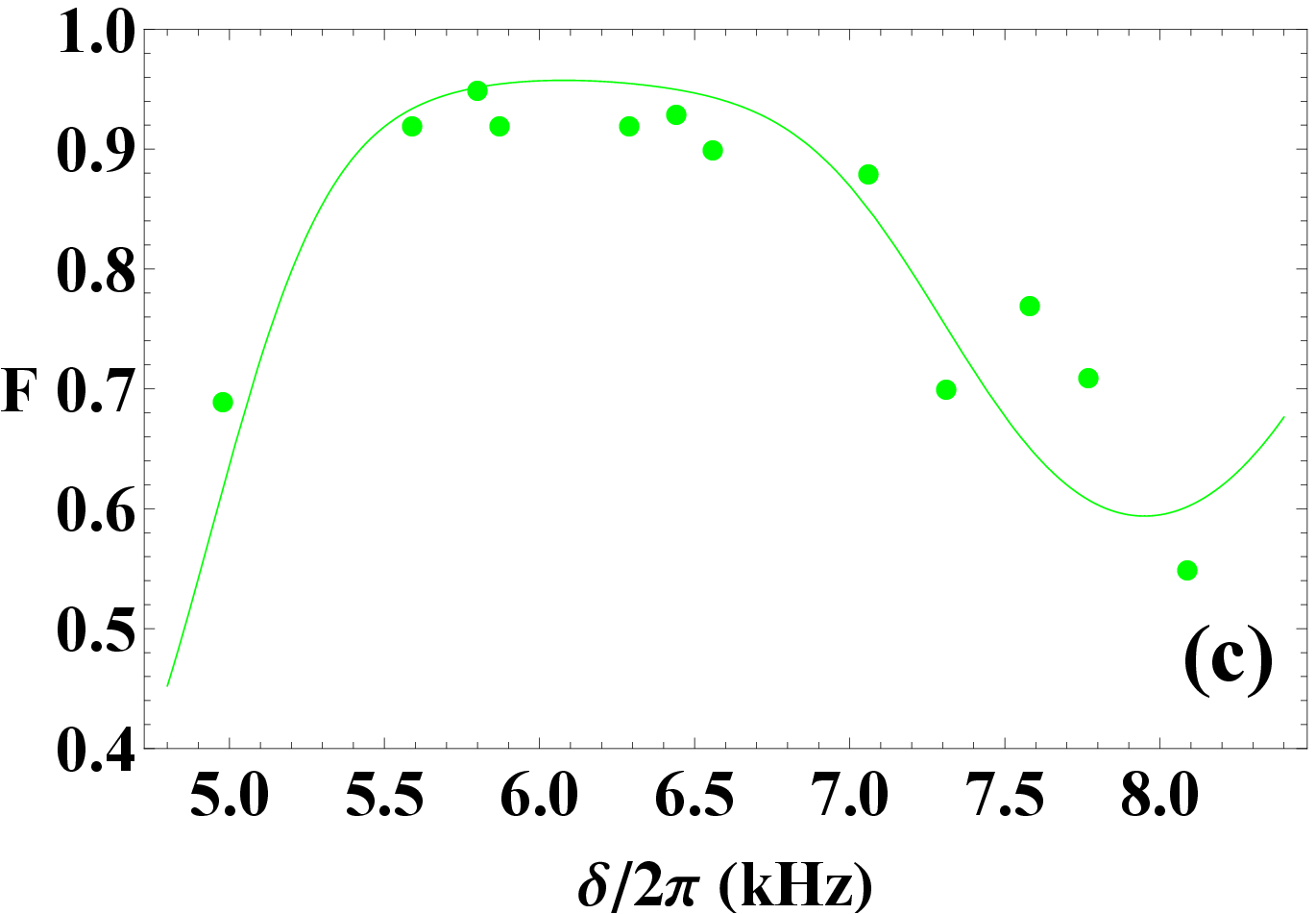}
\includegraphics[width=1.6in]{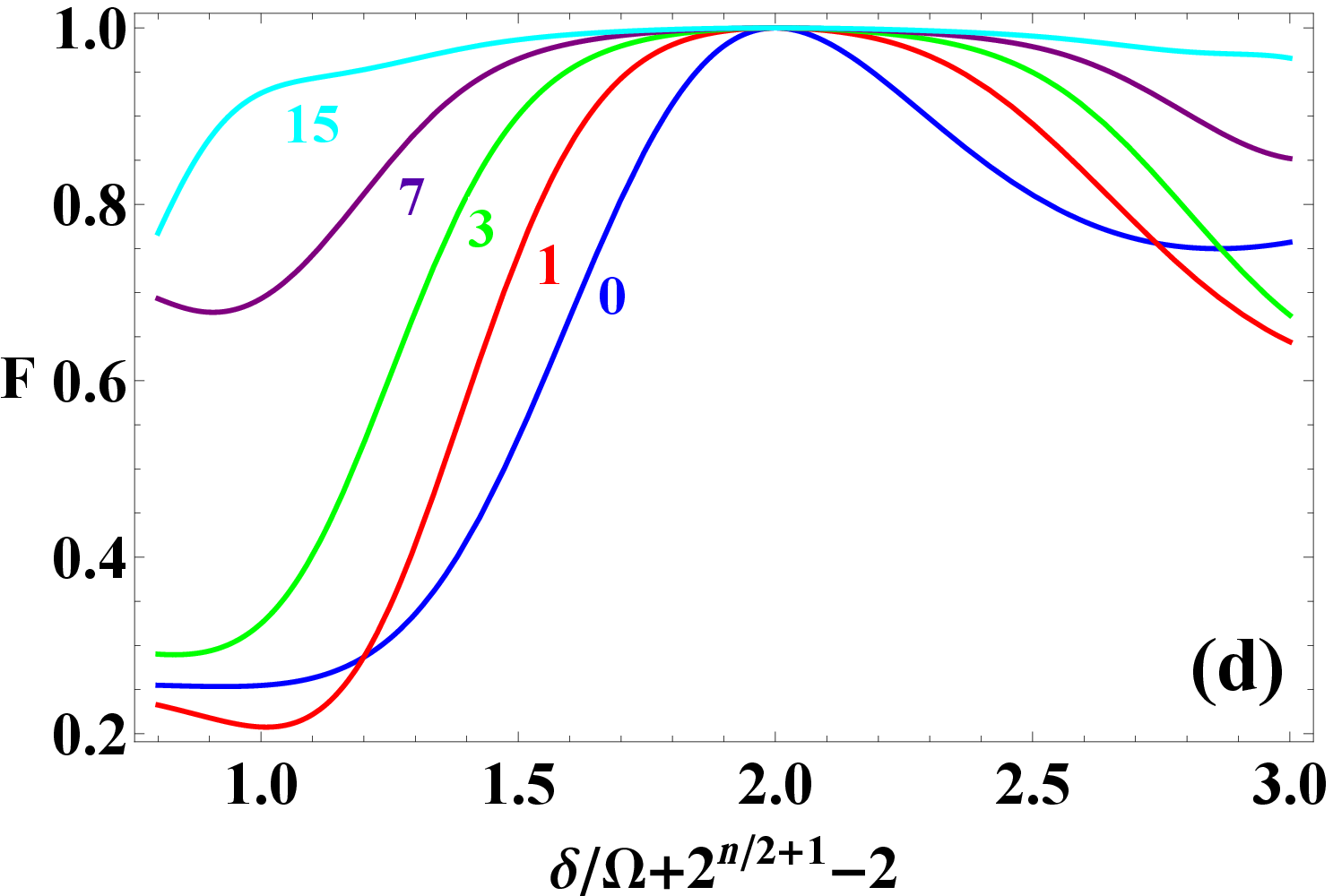}
\caption{The state fidelity of a two ion MS gate as a function of the detuning $\delta$ is compared for the first three Walsh functions being used for $\phi_s$.  In all three data sets, the ions were sideband cooled to the motional ground state before implementing the gate and $t_g=0.3$ms.  In figures (a), (b) and (c) the measured state fidelity as a function of the detuning $\delta$ is compared with the theoretical curves.  To account for other imperfections in the experiment, the theoretical curves have an overall scale factor.  The data shows a maximum fidelity of $\sim0.88$ for the standard pulse, $\sim0.90$ for the $\mathrm{W}(1,t/\sqrt{2}t_g)$ sequence and $\sim0.93$ for $\mathrm{W}(3,t/2t_g)$.  The characteristic width of the high fidelity region is also clearly larger for the higher order pulse sequences as predicted by the theory.  (d) shows the fidelity curves for $\mathrm{W}(0)\mathrm{(blue)},\mathrm{W}(1)\mathrm{(red)},\mathrm{W}(3)\mathrm{(green)},\mathrm{W}(7)\mathrm{(purple)}$ and $\mathrm{W}(15)\mathrm{(cyan)}$.  The curves are shifted in frequency in order to facilitate comparison and highlight the increasingly large regions of high fidelity for the higher order sequences.}
\label{fig:two:ions}
\end{figure}

To demonstrate the power of the composite pulse sequence, we use a qubit defined as the clock states in the $S_{1/2}$ hyperfine manifold of a $\mathrm{Yb}^{+}$ ion in an RF Paul trap which can be initialized and read out using the techniques described in \cite{olmschenk:2007}.  These states, $\left\{\ket{F=0,m_F=0}\equiv\ket{\downarrow}, \ket{F=1,m_F=0}\equiv\ket{\uparrow}\right\}$, have a splitting of $12.6428$ GHz and can be coupled to each other directly or through the harmonic oscillator using stimulated Raman transitions.  As described in \cite{hayes:2010}, the Raman transition induced spin-dependent forces are created by the beat notes between two optical frequency combs that are generated by a $355$nm mode-locked pulsed laser.  The UV pulses have a duration of $\sim10$ psec at a repetition rate of 80.57 MHz and have a center wavelength that is optimal for minimizing off-resonant scattering from the excited $\mathrm{P}$ states in $\mathrm{Yb}^{+}$\cite{wes:2010}.  At the position of the ion, the two beams are cross-polarized and mutually orthogonal to a magnetic field of $5$ G with a geometry such that the momentum kicks associated with the stimulated Raman process only excite the transverse modes of motion which have a resonance frequency of $1.5$ MHz.  The two Raman beams are frequency shifted with AOMs to set up the appropriate beat notes in the interference field at the location of the ions.  As described in \cite{hayes:2010}, driving one AOM with a single frequency and the other with two frequencies generates the bichromatic beat note that gives rise to the MS interaction.  The red and blue phases $\phi_{r/b}$ are therefore defined by the phases of these two RF drive frequencies.  The composite pulse is implemented by splitting the operation into segments, between which the phases $\phi_r$ and $\phi_b$ are shifted.  In this setup, symmetric detuning errors might be the result of fluctuating RF trap voltages which manifests itself as noise on the oscillation frequency. Asymmetric detuning errors arise from a change in the qubit splitting which might stem from a noisy local magnetic field.

The effect of Walsh modulation on the spin-dependent force can be plainly seen with a single ion.  In the case of a single ion, the phase $\Phi(t)$ is global and the only operation that results in a pure spin state is one that simply restores the initial spin state.  The disentanglement of the spin and motion and consequential revival of the initial spin state should occur when $\delta t_g/2\pi=2^nj$ where $2^n-1=k$ is the Walsh function index.  If the spin is initialized to $\ket{\downarrow}$ in the $\hat{\sigma}_z$ basis and the joint spin-motion state after the operation is $\hat{\rho}$, then the state-fidelity is $F_1=\mathrm{Tr}\left[\ket{\downarrow}\bra{\downarrow}\hat{\rho}\right]$.  Ignoring heating effects and assuming an initial thermal state of motion, the fidelity is $F=\frac{1}{2}\left(1+\mathrm{exp}\left[-(\bar{n}+1/2)\left|2\alpha_k(t_g)\right|^2\right]\right)$ where $\bar{n}$ is the average excitation number of the harmonic oscillator and $\alpha_k(t_g)=\frac{\Omega}{2}\int_0^{t_g}dt\mathrm{W}\left(k,t/t_g\right)e^{-i(\delta+\Delta)t}$.  Because the lowest order term for the infidelity is $\mathcal{O}(\left|\alpha_k\right|^2)$, the infidelity of the Walsh modulated operation at $\delta t_g/2\pi=2^nj$ is $\mathcal{O}(\Delta^{2n+2})$.  This effect is clearly seen in Fig. \ref{fig:one:ion} where the higher order Walsh sequences exhibit spin revivals of high purity over a much larger range of detunings.  The increase in the width of the spin revival regions for the higher order Walsh sequences is evidence of the higher tolerance of small detuning errors.  Note the same arguments apply to the case of small timing errors and errors resulting from an asymmetric detuning from the red and blue sidebands such as a change in the qubit splitting.

The effect of the Walsh modulation on a two-qubit gate is more complicated than that of a single qubit operation since the term in (\ref{eq:time:evolution}) proportional to $\hat{S}^2_N$ must be taken into account.  The Walsh modulation of $\phi_s$ changes the evolution of $\Phi(t)$ in general, but not in the case where $\delta=2^{n+1}\pi/t_g$ since the evolution is a series of closed circles in phase space.  In this case, $\Phi(t_g)=\Omega^2t_g/\delta$ and a fully entangling operation is achieved when $\Phi(t_g)=\pi/2$.  This implies that in order to use $W\left(2^n-1,t/t_g\right)$, the gate time must be at least $t_g=2^{n/2}\pi/\Omega$.  While the exponential nature of this composite gate becomes daunting for large $n$, small errors can easily be corrected with a modest increase in the gate time.  In the case of two ions, the maximally entangling operation ideally implements the transformation $\ket{\downarrow\downarrow}\Rightarrow\ket{\downarrow\downarrow}+e^{i\theta}\ket{\uparrow\uparrow}$ where the phase $\theta$ is determined by the phase of the drive field.  With this target state, the fidelity is $F_2=1/4\left|e^{-\left(\bar{n}+1/2\right)\left|2\alpha_k(t_g)\right|^2}+ie^{-i\Omega^2\Phi_k(t_g)}\right|^2$ and is measured in the same manner as described in \cite{hayes:2010}.  The phase $\Phi_k(t_g)=\sum_{i>j=0}^k\mathrm{Im}\left[\varphi^*_i\varphi_j\right]-\frac{1}{\delta}\left(t_g-\frac{1}{\delta}\sum_{i=0}^{k}\mathrm{sin}(\delta t_i)\right)$ is written here in terms of sums over the different parts of a pulse sequence.  The parameters $t_i$ refer to the duration of the $(i+1)^{th}$ segment of a sequence and the parameters $\varphi_i=(-)^i\int_{t_{i-1}}^{t_i}dte^{-i\delta t}$ with $t_{-1}=0$ and $t_k=t_g$.  The data shown in Fig. \ref{fig:two:ions} compare the state fidelity of two ions using the composite pulse sequences for the maximally entangling gate and shows that the standard pulse sequence is outperformed by the Walsh modulated pulse sequences in terms of both the maximum fidelity and the characteristic width of the high fidelity region.  The value of $\Omega$ is used as a fit parameter for the theoretical curves with the same value, $2\pi\times1.47$ kHz, used in all three plots.

Walsh functions have long been known by the electrical engineering, astronomy and radio communications communities to have useful error correcting properties \cite{beauchamp:1984}.  While the Walsh functions are not the only option for choosing how to modulate the drive field of the spin-dependent force gate, we hope their introduction in the context of quantum control provides a useful tool for the further development of dynamical decoupling and related areas.  In the formalism of dynamical decoupling, the function $\alpha_k(t_g)$ can be viewed as an optimized filter function designed to suppress the effects of a noise source centered at $\delta/2\pi$ \cite{cywinsky:2008}.  Walsh modulation is optimal in a different sense than that of the Uhrig sequence which can, with an increasing number of pulses, suppresses higher moments of the noise spectrum in the zero frequency limit \cite{uhrig:2007}.  While the Uhrig filter is optimal in the number of pulses used for a given order of noise suppression, the Walsh filters are optimal in the number of elementary sequences, (see supplemental material for the definition of an elementary sequence).  This minimal number of elementary sequences not only allows for a simple mathematical construction, but also means that these functions are easy to synthesize using simple integrated circuits.

By introducing the idea of a Walsh modulated spin-dependent force, we have shown theoretically and experimentally that it is possible to suppress errors that are linked to the residual entanglement between the spin and motion, thereby alleviating the required precision of the control fields.  Because the detrimental effect of any error increases exponentially with the initial temperature of the harmonic oscillator, the technique may also decrease the amount of resource intensive cooling that must be done in order to achieve a high fidelity operation.  As quantum information experiments progress, this technique of coherent error suppression in quantum bus operations might prove to be an important ingredient in scaling toward larger systems and more complex algorithms. 
\begin{acknowledgments}
We acknowledge useful discussions with Peter Maunz, Daniel Brennan and Michael Biercuk.  This work is supported by the Army Research Office (ARO) with funds 
from the DARPA Optical Lattice Emulator (OLE) Program, IARPA under 
ARO contract, the NSF Physics at the Information Frontier Program, 
and the NSF Physics Frontier Center at JQI.
\end{acknowledgments}
\bibliographystyle{prl}

\clearpage
\appendix
\section{Supplemental Material}
In this section, we aim to prove the following identity by induction,
\begin{equation}
\int_0^1dx\mathrm{W}(2^n-1,x)e^{i2^{n+1}\pi x}\sum_{l=0}^na_lx^l=0.
\label{eq:prove}
\end{equation}
The construction of the Walsh functions is simple in terms of the elementary sequences known as Rademacher functions $\mathrm{R}\left(n,t\right)=\mathrm{sign}\left[\mathrm{sin}\left(2^n\pi t\right)\right]$.  The dyadic ordering of the Walsh functions allow them to be defined in terms of the Rademacher functions as $\mathrm{W}\left(n,t\right)=\prod_{i=1}^{m+1}\mathrm{R}\left(i,t\right)^{b_{i-1}}$ when $n$ is expressed as a binary number $n=b_m2^m+....+b_02^0$ and $b_i=0$ or $1$.  With this definition, it is easy to see that choosing the index $2^n-1$ for the Walsh function means that all the $b_i$ coefficients are $1$.  We now prove the base case, $n=1$.
\begin{eqnarray*}
& &\int_0^1dx\mathrm{W}(1,x)e^{i4\pi x}\sum_{l=0}^1a_lx^l\\
&=&\int_0^1dx\mathrm{R}(1,x)e^{i4\pi x}\sum_{l=0}^1a_lx^l\\
&=&a_1\int_0^1dx\mathrm{R}(1,x)e^{i4\pi x}x\\
&=&a_1\left(\int_0^{1/2}dxe^{i4\pi x}x-\int_{1/2}^1dxe^{i4\pi x}x\right)\\
&=&a_1\int_0^{1/2}dxe^{i4\pi x}\left(x-(x+1/2)\right)=0.
\label{eq:base:case}
\end{eqnarray*}
For the inductive step, assume that (\ref{eq:walsh:coefficients}) is true and look at the $2^{n+1}-1$ case:
\begin{eqnarray*}
& &\int_0^1dx\mathrm{W}(2^{n+1}-1,x)e^{i2^{n+2}\pi x}\sum_{l=0}^{n+1}a_lx^{l}\\
&=&\int_0^1dx\mathrm{R}(1,x)....\mathrm{R}(n+1,x)e^{i2^{n+2}\pi x}\sum_{l=0}^{n+1}a_lx^{l}\\
&=&\frac{1}{2}\int_{0}^{2}dx\prod_{i=1}^{n+1}\mathrm{R}(i,x/2)e^{i2^{n+1}\pi x}\sum_{l=0}^{n+1}a_l\left(\frac{x}{2}\right)^{l}\\
&=&\frac{1}{2}\int_{0}^{1}dx\prod_{i=1}^{n+1}\mathrm{R}(i,x/2)e^{i2^{n+1}\pi x}\sum_{l=0}^{n+1}a_l\left(\frac{x}{2}\right)^{l}\\
&+&\frac{1}{2}\int_{1}^{2}dx\prod_{i=1}^{n+1}\mathrm{R}(i,x/2)e^{i2^{n+1}\pi x}\sum_{l=0}^{n+1}a_l\left(\frac{x}{2}\right)^{l}.
\label{eq:inductive:step:1}
\end{eqnarray*}
For the next step, note that for an integer $n\geq0$, $\mathrm{R}(n+1,x/2)=\mathrm{R}(n,x)$, which allows the expression to be written as,
\begin{eqnarray*}
&=&\frac{1}{2}\int_{0}^{1}dx\prod_{i=0}^{n}\mathrm{R}(i,x)e^{i2^{n+1}\pi x}\sum_{l=0}^{n+1}a_l\left(\frac{x}{2}\right)^{l}\\
&+&\frac{1}{2}\int_{1}^{2}dx\prod_{i=0}^{n}\mathrm{R}(i,x)e^{i2^{n+1}\pi x}\sum_{l=0}^{n+1}a_l\left(\frac{x}{2}\right)^{l}\\
&=&\frac{1}{2}\int_{0}^{1}dx\prod_{i=1}^{n}\mathrm{R}(i,x)e^{i2^{n+1}\pi x}\sum_{l=0}^{n+1}a_l\left(\frac{x}{2}\right)^{l}\\
&-&\frac{1}{2}\int_{1}^{2}dx\prod_{i=1}^{n}\mathrm{R}(i,x)e^{i2^{n+1}\pi x}\sum_{l=0}^{n+1}a_l\left(\frac{x}{2}\right)^{l}.
\label{eq:inductive:step:2}
\end{eqnarray*}
In the next step, the substitution $x'=x-1$ is made and we take advantage of the fact that $\mathrm{R}(n,x+1)= \mathrm{R}(n,x)$ for $n\geq1$.
\begin{eqnarray*}
&=&\frac{1}{2}\int_{0}^{1}dx\mathrm{W}(2^n-1,x)e^{i2^{n+1}\pi x}\sum_{l=0}^{n+1}\frac{a_l}{2^l}x^{l}\\
&-&\frac{1}{2}\int_{0}^{1}dx\mathrm{W}(2^n-1,x)e^{i2^{n+1}\pi x}\sum_{l=0}^{n+1}\frac{a_l}{2^l}\sum_{k=0}^l\binom{m}{k}x^k\\
&=&\frac{1}{2}\int_{0}^{1}dx\mathrm{W}(2^n-1,x)e^{i2^{n+1}\pi x}\sum_{l=0}^{n}b_lx^l,
\label{eq:inductive:step:3}
\end{eqnarray*}
which is zero by assumption since $b_l$ is a constant, thus concluding the proof.  If the Radamacher functions are considered to be elementary sequences, then the Walsh filter is optimized in this resource for the task of suppressing the errors in the spin-dependent force operation that are discussed in this paper.
\end{document}